\documentclass{elsart}
\usepackage{times}
\usepackage{courier}
\usepackage{algorithm}
\usepackage{algorithmic}
\usepackage{natbib}
\usepackage{epsfig}
\usepackage{amssymb}

\def\dj{\hbox{d\kern-0,347em \vrule width0,3em height1,252ex
depth-1,21ex \kern0,051em}}
\begin{document}

\begin{frontmatter}

\title{An Efficient Block Circulant Preconditioner For Simulating Fracture Using Large Fuse Networks}

\author[1]{Phani Kumar V.V. Nukala}
\author[1]{Sr{\dj}an \v{S}imunovi\'{c}}
\address[1]{Computer Science and Mathematics Division, 
Oak Ridge National Laboratory, Oak Ridge, TN 37831-6164, USA}

\begin{abstract}
{\it Critical slowing down} associated with the iterative solvers close
to the critical point often hinders large-scale numerical simulation 
of fracture using discrete lattice networks. This paper presents a block circlant 
preconditioner for iterative solvers for the 
simulation of progressive fracture in disordered, quasi-brittle materials using large discrete lattice
networks. The average computational cost of the present alorithm per iteration is 
$O(rs~log ~s) + delops$, where the stiffness matrix ${\bf A}$ is partioned into 
$r$-by-$r$ blocks such that each block is 
an $s$-by-$s$ matrix, and $delops$ represents the operational count associated with 
solving a block-diagonal matrix with $r$-by-$r$ dense matrix blocks. 
This algorithm using the block circulant preconditioner 
is faster than the Fourier accelerated preconditioned conjugate gradient (PCG) 
algorithm, and alleviates 
the {\it critical slowing down} that is especially severe close to the critical point.
Numerical results using random resistor networks substantiate
the efficiency of the present algorithm.
\end{abstract}


\end{frontmatter}

\section{Introduction}
Progressive damage evolution leading to failure of disordered
quasi-brittle materials has been
studied extensively using various types of discrete lattice models
\cite{deArcangelis85,sahimi86,duxbury86,duxbury87,hansen001,herrmann90,sahimi98,chakrabarti}.
Numerical simulation of large lattice networks has often been hampered due to 
{\it critical slowing down} associated with the iterative solvers as the lattice 
system approaches macroscopic fracture. The authors have developed a 
multiple-rank sparse Cholesky update algorithm based on direct solvers for simulating fracture using 
discrete lattice systems \cite{nukalajpamg}. Using the algorithm presented in 
\cite{nukalajpamg}, the authors have reported numerical simulation results for large 
2D lattice systems (e.g., $L = 512$), which to the authors knowledge, was so far the 
largest lattice system used in studying damage evolution using discrete lattice systems. 
Although the sparse direct 
solvers presented in \cite{nukalajpamg} are superior to iterative solvers in two-dimensional 
lattice systems, for 3D lattice systems, the memory demands brought about by 
the amount of {\it fill-in} during sparse {\it Cholesky} factorization 
favor iterative solvers. Hence, iterative solvers are 
in common use for large-scale 3D lattice simulations. 
As the lattice system gets closer to macroscopic fracture, the condition
number of the system of linear equations increases, thereby increasing
the number of iterations required to attain a fixed accuracy.  This
becomes particularly significant for large lattices.  
Fourier accelerated PCG iterative
solvers \cite{batrouni86,batrouni88,batrouni98} have been used in the
past to alleviate the critical slowing down. 
However, the Fourier acceleration technique based on ensemble averaged 
circulant preconditioner is not effective when
fracture simulation is performed using central-force and bond-bending
lattice models \cite{batrouni88}. The main focus of the current paper is 
on developing an efficient algorithm based on iterative solvers 
for large-scale 3D lattice simulations, and the 
block-circulant preconditioner presented in the current paper is an effort 
towards this goal. 

\par
Since the Laplacian operator on a 
discrete lattice network results in the block structure of the stiffness matrix, 
we propose to use block circulant matrices \cite{chan92,chan96} as preconditioners to the stiffness matrix 
for solving this class of problems.
The proposed algorithm is benchmarked against the commonly used 
incomplete LU and Cholesky preconditioners \cite{golub96}, and the 
{\it optimal} \cite{tchan88,chan89,chan921,chan96} 
and {\it superoptimal} \cite{tyrty,chan96} circulant preconditioners  
to the Laplacian operator (Kirchhoff equations).
The advantage of using the 
circulant preconditioners is that they can be diagonalized by discrete Fourier matrices, and 
hence the inversion of $n_{dof}$-by-$n_{dof}$ circulant matrix can be done in 
$O(n_{dof} ~log ~n_{dof})$ operations by using FFTs of size $n_{dof}$. 
In addition, since the convergence rate of the PCG method depends on the
condition number of the preconditioned system, it is possible to 
choose a circulant preconditioner that minimizes the condition number of 
the preconditioned system \cite{tyrty,chan96}. Furthermore, these circulant 
preconditioned systems exhibit favorable clustering of eigenvalues. In 
general, the more clustered the eigenvalues are, the faster the 
convergence rate is. Another important property of these circulant preconditioners 
proposed in this study is that they are positive definite if the stiffness matrix itself is positive definite. 
In this regard, 
we note that the Fourier accelerated PCG presented in \cite{batrouni86,batrouni88,batrouni98}
is not optimal in the sense described in \cite{tchan88,chan89,chan921,chan96}, and 
hence is expected to take more number of CG iterations compared with the {\it optimal} and 
{\it superoptimal} circulant preconditioners.

\par
In this paper, we analyze a {\it random threshold} model problem, where a lattice consists of 
fuses having the same conductance, but the bond breaking thresholds, $i_c$, 
are based on a broad (uniform) probability distribution, which is constant between
0 and 1. This relatively simple model has been extensively used in the literature for 
simulating the fracture and progressive damage evolution in brittle materials, and provides a 
meaningful benchmark for comparing different algorithms. 
A broad thresholds distribution represents large disorder and
exhibits diffusive damage leading to progressive localization, whereas a very narrow
thresholds distribution exhibits brittle failure in which a single crack propagation
causes material failure. Periodic boundary conditions are imposed in the horizontal direction to simulate
an infinite system and a constant voltage difference (displacement)
is applied between the top and
the bottom of lattice system. The simulation is initiated with a triangular lattice of intact
fuses of size $L \times L$, in which disorder is introduced through random breaking thresholds. The voltage
$V$ across the lattice system is increased until a fuse (bond breaking) burns out.
The burning of a fuse occurs whenever the electrical current (stress)
in the fuse (bond) exceeds the
breaking threshold current (stress) value of the fuse. The current is redistributed
instantaneously after a fuse is burnt. The voltage is then gradually increased until
a second fuse is burnt, and the process is repeated.
Each time a fuse is removed, the electrical current is redistributed and hence it is 
necessary to re-solve Kirchhoff equations to determine the current flowing in the
remaining bonds of the lattice. This step is essential for determining 
the fuse that is going to burn up under the redistributed currents. Therefore, 
numerical simulations leading to final breaking of lattice system network are very 
time consuming especially with increasing lattice system size. Consequently, an efficient 
preconditioner to the Laplacian operator on fractal networks that mitigates the effect of 
critical slowing down as the lattice system approaches macroscopic fracture is of utmost 
importance in the numerical simualtion of material breakdown.

\par
In the following, we present point-circulant and block circulant preconditioners for 
solving the linear system of equations that arise during the numerical simulation of 
progressive fracture in brittle materials using the random threshold model.

\section{Circulant Preconditioners for CG Iterative Solvers}
Consider the $n_{dof} \times n_{dof}$ stiffness matrix ${\bf A}$. The {\it optimal} circulant 
preconditioner $c({\bf A})$ \cite{tchan88} is defined as the minimizer of 
$\|{\bf C} - {\bf A}\|_F$ over all $n_{dof} \times n_{dof}$ circulant matrices ${\bf C}$. 
In the above description, $\| \cdot \|_F$ denotes the Frobenius norm \cite{golub96}, 
and the matrix $c({\bf A})$ 
is called an {\it optimal} circulant preconditioner because it minimizes the 
norm $\|{\bf C} - {\bf A}\|_F$. The {\it optimal} circulant preconditioner $c({\bf A})$ is 
uniquely determined by ${\bf A}$, and is given by
\begin{eqnarray}
c({\bf A}) & = & {\bf F}^{\ast} \delta\left({\bf F} {\bf A} {\bf F}^{\ast}\right) {\bf F}
\end{eqnarray}
where ${\bf F}$ denotes the discrete Fourier matrix, $\delta\left({\bf A}\right)$ denotes the 
diagonal matrix whose diagonal is equal to the diagonal of the matrix ${\bf A}$, and 
$\ast$ denotes the adjoint (i.e. conjugate transpose). It should be noted that the diagonal 
elements of the matrix $\delta\left({\bf F} {\bf A} {\bf F}^{\ast}\right)$ represent the 
eigenvalues of the matrix $c({\bf A})$ and can be obtained in $O(n_{dof} ~log ~n_{dof})$ 
operations by taking the FFT of the first column of $c({\bf A})$. The first 
column vector of T. Chan's {\it optimal} 
circulant preconditioner matrix that minimizes the norm $\|{\bf C} - {\bf A}\|_F$ is given by
\begin{eqnarray}
c_i & = & \frac{1}{n_{dof}} \sum_{j=1}^{n_{dof}} a_{j,(j-i+1)~\mbox{mod}~n_{dof}} \label{tchaneq}
\end{eqnarray}
The above formula can be interpreted simply as follows: the element $c_i$ is simply the 
arithmetic average of those diagonal elements of ${\bf A}$ extended to length 
$n_{dof}$ by wrapping around and containing the element $a_{i,1}$. If the matrix ${\bf A}$ is 
a Hermitian matrix, then the eigenvalues of $c({\bf A})$ are bounded from below and above by 
\begin{equation}
\lambda_{min}({\bf A}) \leq \lambda_{min}(c({\bf A})) \leq \lambda_{max}(c({\bf A})) \leq \lambda_{max}({\bf A}) \label{pspect}
\end{equation}
where $\lambda_{min}(\cdot)$ and $\lambda_{max}(\cdot)$ denote the minimum and maximum eigenvalues, 
respectively. Based on the above result, if the matrix ${\bf A}$ is positive definite, then 
the circulant preconditioner $c({\bf A})$ is also positive definite. In particular, if the 
circulant preconditioner is such that the spectra of the preconditioned system is 
clustered around $1$, then the convergence of the solution will be fast. The 
{\it superoptimal} circulant preconditioner $t({\bf A})$ \cite{tyrty} is based on the idea of 
minimizing the norm $\|{\bf I} - {\bf C}^{-1} {\bf A}\|_F$ over all nonsingular 
circulant matrices ${\bf C}$. In the above description, $t({\bf A})$ is {\it superoptimal} 
in the sense that it minimizes $\|{\bf I} - {\bf C}^{-1} {\bf A}\|_F$, and is equal to
\begin{equation}
t({\bf A}) = c({\bf A} {\bf A}^{\ast}) c({\bf A})^{-1} \label{tyrtyeq}
\end{equation}
The preconditioner obtained by Eq. (\ref{tyrtyeq}) is also positive definite if the matrix 
${\bf A}$ itself is positive definite. Although the preconditioner $t({\bf A})$ is 
obtained by minimizing the norm $\|{\bf I} - {\bf C}^{-1} {\bf A}\|_F$, the asymptotic 
convergence of the preconditioned system is same as $c({\bf A})$ for large $n_{dof}$ system.
Hence, in this study, we limit ourselves to the investigation of preconditioned systems 
using $c({\bf A})$ given by Eq. (\ref{tchaneq}). The computational cost associated with 
the solution of preconditioned system $c({\bf A}) {\bf z} = {\bf r}$ is the initialization cost of 
$nnz({\bf A})$ for setting the first column of $c({\bf A})$ using Eq. (\ref{tchaneq}) 
during the first iteration, and $O(n_{dof} ~log ~n_{dof})$ during every iteration step.

\par
In order to distinguish the block circulant preconditioners that follow from the 
above described circulant preconditioners, we refer henceforth to the above 
preconditioners as point-circulant preconditioners.

\subsection{Block-circulant preconditioners}
Let the matrix ${\bf A}$ is partioned into $r$-by-$r$ blocks such that each block is 
an $s$-by-$s$ matrix. That is, $n_{dof} = rs$, and 
\begin{eqnarray}
{\bf A} & = & \left[\begin{array}{cccccccc}
{\bf A}_{1,1} & {\bf A}_{1,2} & \cdots & {\bf A}_{1,r} \\
{\bf A}_{2,1} & {\bf A}_{2,2} & \cdots & {\bf A}_{2,r} \\
\vdots & \vdots & \ddots & \vdots \\
{\bf A}_{r,1} & {\bf A}_{r,2} & \cdots & {\bf A}_{r,r} \\
\end{array} \right] \label{Abb}
\end{eqnarray}
Although the point-circulant preconditioner $c({\bf A})$ defined by Eq. (\ref{tchaneq}) can 
be used as a preconditioner, in general, the block structure is not restored by 
using $c({\bf A})$ as a preconditioner. In contrast, the circulant-block preconditioners 
obtained by using circulant approximations for each of the blocks restore the 
block structure of ${\bf A}$. The circulant-block preconditioner of ${\bf A}$ 
can be expressed as
\begin{eqnarray}
c_B({\bf A}) & = & \left[\begin{array}{cccccccc}
c({\bf A}_{1,1}) & c({\bf A}_{1,2}) & \cdots & c({\bf A}_{1,r}) \\
c({\bf A}_{2,1}) & c({\bf A}_{2,2}) & \cdots & c({\bf A}_{2,r}) \\
\vdots & \vdots & \ddots & \vdots \\
c({\bf A}_{r,1}) & c({\bf A}_{r,2}) & \cdots & c({\bf A}_{r,r}) \\
\end{array} \right] \label{cAbb}  
\end{eqnarray}
The circulant-block preconditioner defined by Eq. (\ref{cAbb}) is the minimizer of 
$\|{\bf C} - {\bf A}\|_F$ over all matrices ${\bf C}$ that are $r$-by-$r$ block matrices 
with $s$-by-$s$ circulant blocks.
The spectral properties as given by Eq. (\ref{pspect}) for point-circulant preconditioners 
also extend to the circulant-block preconditioners \cite{chan92,chan96}. That is, 
\begin{equation}
\lambda_{min}({\bf A}) \leq \lambda_{min}(c_B({\bf A})) \leq \lambda_{max}(c_B({\bf A})) \leq \lambda_{max}({\bf A}) \label{bspect}
\end{equation}
In particular, if the matrix ${\bf A}$ is positive definite, then the block-preconditioner 
$c_B({\bf A})$ is also positive definite.

\par
The computational cost associated with the circulant-block preconditioners can be estimated as 
follows. Since the stiffness matrix ${\bf A}$ is real symmetric for the type of problems 
considered in this study, in the following, we assume block symmetric structure for ${\bf A}$,
i.e., ${\bf A}_{j,i} = {\bf A}_{i,j}^t$. In forming the circulant-block preconditioner 
given by Eq. (\ref{cAbb}), it is necessary to obtain point-circulant preconditioners for 
each of the $r$-by-$r$ block matrices of order $s$. Point-circulant approximation for 
each of the $s$-by-$s$ blocks requires $O(s ~log ~s)$ operations. This cost is 
in addition to the cost associated in forming the first column vectors (Eq. (\ref{tchaneq})) 
for each of the 
$c({\bf A}_{i,j})$ blocks, which is given by $nnz({\bf A})$ operations. Since there are 
$(r(r+1))/2$ blocks, we need 
$O(r^2 s ~log ~s)$ operations to form
\begin{eqnarray}
{\bf \Delta} & = & ({\bf I} \otimes {\bf F}) c_B({\bf A}) ({\bf I} \otimes {\bf F}^{\ast}) 
= \left[\begin{array}{cccccccccc}
\delta\left({\bf F} {\bf A}_{1,1} {\bf F}^{\ast}\right) & 
\delta\left({\bf F} {\bf A}_{1,2} {\bf F}^{\ast}\right) & \cdots & 
\delta\left({\bf F} {\bf A}_{1,r} {\bf F}^{\ast}\right) \\
\delta\left({\bf F} {\bf A}_{2,1} {\bf F}^{\ast}\right) & 
\delta\left({\bf F} {\bf A}_{2,2} {\bf F}^{\ast}\right) & \cdots & 
\delta\left({\bf F} {\bf A}_{2,r} {\bf F}^{\ast}\right) \\
\vdots & \vdots & \ddots & \vdots \\
\delta\left({\bf F} {\bf A}_{r,1} {\bf F}^{\ast}\right) & 
\delta\left({\bf F} {\bf A}_{r,2} {\bf F}^{\ast}\right) & \cdots & 
\delta\left({\bf F} {\bf A}_{r,r} {\bf F}^{\ast}\right) \\
\end{array} \right] \label{Del}
\end{eqnarray}
In the above equation, $\otimes$ refers to the Kronecker tensor product and ${\bf I}$ is an 
$r$-by-$r$ identity matrix. In order to solve the preconditioned 
equation $c_B({\bf A}) {\bf z} = {\bf r}$, the Eq. (\ref{Del}) is permuted to obtain a 
block-diagonal matrix of the form
\begin{eqnarray}
\tilde{{\bf \Delta}} & = & {\bf P}^{\ast} {\bf \Delta} {\bf P} = \left[\begin{array}{cccccccccc}
\tilde{{\bf \Delta}}_{1,1} & {\bf 0} & \cdots & {\bf 0} \\
{\bf 0} & \tilde{{\bf \Delta}}_{2,2} & \cdots & {\bf 0} \\
\vdots & \vdots & \ddots & \vdots \\
{\bf 0} & {\bf 0} & \cdots & \tilde{{\bf \Delta}}_{s,s} \\
\end{array} \right] \label{Delper}
\end{eqnarray}
where ${\bf P}$ is the permutation matrix such that
\begin{eqnarray}
\left[\tilde{{\bf \Delta}}_{k,k}\right]_{ij} & = & 
\left[\delta\left({\bf F} {\bf A}_{i,j} {\bf F}^{\ast}\right)\right]_{kk}
~~~\forall ~1 \leq i, j \leq r, ~~1 \leq k \leq s
\end{eqnarray}
During each iteration step, in order to solve the preconditioned system 
$c_B({\bf A}) {\bf z} = {\bf r}$, it is necessary to invert the block-diagonal matrix 
$\tilde{{\bf \Delta}}$. This task can be performed by first factorizing each of the 
$\tilde{{\bf \Delta}}_{k,k}$ blocks during the first iteration step, and then subsequently 
using these factored matrices to do the baclsolve operations. Hence, without considering the 
first factorizing cost of each of the block diagonals, during each 
iteration step, the number of operations involving the 
inversion of $\tilde{{\bf \Delta}}$ is  
\begin{eqnarray}
delops & = & O(\sum_{k=1}^{s} |{\mathcal L}_{\tilde{{\bf \Delta}}_{k,k}}|) \label{delops}
\end{eqnarray}
where ${\mathcal L}_{\tilde{{\bf \Delta}}_{k,k}}$ denotes the number of non-zeros in the 
Cholesky factorization of the matrix ${\tilde{{\bf \Delta}}_{k,k}}$. Therefore, the system 
$c_B({\bf A}) {\bf z} = {\bf r}$ can be solved in $O(rs~log ~s) + delops$ operations 
per iteration step. Thus, we conclude that for the circulant-block 
preconditioner, the initialization cost is $nnz({\bf A}) + O(r^2 s ~log ~s)$ plus the cost 
associated with the factorization of each of the diagonal blocks $\tilde{{\bf \Delta}}_{k,k}$ 
during the first iteration, and $O(rs~log ~s) + delops$ during every iteration step.

\par
Although from operational cost per iteration point of view, the point-circulant 
preconditioner may prove advantageous for some problems, it is not clear whether 
point-circulant or circulant-block is closest to the matrix ${\bf A}$ in terms of 
the number of CG iterations necessary for convergence. Hence, we investigate both 
point-circulant and circulant-block preconditioners in obtaining the solution of 
the linear system ${\bf A} {\bf x} = {\bf b}$ using iterative techniques.
In addition, we also employ the commonly used point and block versions of 
the {\it incomplete} LU preconditioners to solve the linear system ${\bf A} {\bf x} = {\bf b}$.

\vskip 0.70em%
\noindent
REMARK 1: In the case of 2D discrete lattice network with periodic boundary conditions 
in the horizontal direction and a constant voltage difference between the top and the 
bottom of the lattice network, the matrix ${\bf A}$ is a block tri-diagonal real symmetric 
matrix. Under these circumstances, 
the initialization cost is $nnz({\bf A}) + O(rs ~log ~s)$. Since each of the 
diagonal blocks $\tilde{{\bf \Delta}}_{k,k}$ is a $2 \times 2$ matrix, during 
each iteration step, the solution involving the inversion of $\tilde{{\bf \Delta}}$ 
can be obtained in $O(s)$ operations. Thus, the cost per iteration is 
$O(rs~log ~s) + O(s) = O(rs~log ~s)$ operations. The total computational cost 
involved in using the circulant-block preconditioner for a 
symmetric block tri-diagonal matrix is the initialization cost 
of $nnz({\bf A}) + O(rs ~log ~s)$, and $O(rs~log ~s)$ operations per iteration step.
This is significantly less than the computational cost involved in using a 
generic circulant-block preconditioner. 
It should be noted that the block tri-diagonal structure of ${\bf A}$ does not 
change the computational cost associated with using a point-circulant preconditioner 
to solve the linear system ${\bf A} {\bf x} = {\bf b}$.

\section{Numerical Simulation Results}
In the following, 
we benchmark the proposed block circulant preconditioner against the 
{\it optimal} \cite{tchan88,chan89,chan921,chan96} 
circulant preconditioner used for the Laplacian operator (Kirchhoff equations). 
The main purpose behind the 2D lattice simulations presented below is to demonstrate the 
efficiency of block-circulant preconditioner over the {\it optimal} circulant 
preconditioner for the iterative solvers. 
Once again, we note that the type of ensemble-averaged circulant preconditioner 
presented in \cite{batrouni86,batrouni88,batrouni98}
is not optimal in the sense described in \cite{tchan88,chan89,chan921,chan96}, and 
hence is expected to take more number of CG iterations compared with the {\it optimal} 
circulant preconditioners. In the case of 2D lattice systems, we also present the 
simulation results using {\it Solver type A} of the Ref. \cite{nukalajpamg} based on sparse direct solvers.  
As noted earlier, the sparse direct solvers presented in \cite{nukalajpamg} are superior 
to the iterative solvers for 2D lattice systems, even with the block-circulant preconditioner 
presented in the current paper. However, the main advantage of the block-circulant 
preconditioner using iterative solvers is in the case of simulation of 3D lattice systems, 
where the usage of sparse direct solvers is limited by the (random access) memory constraints.  

\par
The numerical results presented in
Tables 1-5 (for 2D lattices) and 7-10 (for 3D lattices) are performed on a single processor of
{\it Cheetah} (27 Regatta nodes with thirty two 1.3 GHz Power4 processors each, 
{\it http://www.ccs.ornl.gov}). However, the numerical simulation results presented in
Tables 6 (for 2D lattices) and 11 (for 3D lattices) are performed on a single processor of 
{\it Eagle} (184 nodes with four 375 MHz Power3-II processors)
supercomputer at the Oak Ridge National Laboratory to run simulations simultaneously
on more number of processors.
In all of the iterative schemes presented below, we employ a residual tolerance of 
$\epsilon = 10^{-12}$ for convergence of the iterations.
Tables 1 and 2 present the cpu and wall-clock times taken on a single processor of 
{\it Cheetah} for one configuration (simulation) 
using the block circulant and the {\it optimal} circulant precondioned CG iterative solvers, respectively. 
In the case of two-dimensional block circulant PCG, 
we partition the matrix ${\bf A}$ into $L$-by-$L$ blocks such that 
each block is a $(L+1) \times (L+1)$ matrix.
For comparison purposes, we also present in Tables 3 and 4, 
the cpu and wall-clock times taken by un-preconditioned and 
incomplete Cholesky preconditioned CG solvers. Table 5 presents the 
performance of the sparse direct solver ({\it Solver type a}) reported in 
\cite{nukalajpamg}. As discussed earlier, for 2D lattice systems, the sparse direct solvers and the 
incomplete Cholesky preconditioner are clearly superior to the block-circulant 
preconditioned CG iterative solver. However, this advantage of direct solvers (or the 
preconditioners such as incomplete Cholesky based on direct solvers) vanishes for 
large 3D lattice systems due to the amount of {\it fill-in} during 
Cholesky factorization. 
In tables 1-5, $N_{config}$ indicates the 
number of configurations over which ensemble averaging of the 
numerical results is performed, and the  
number of iterations denote 
the average number of total iterations taken to break one intact lattice configuration 
until it falls apart. For some iterative solvers,
the simulations for larger lattice systems were not performed either 
because they were expected to take larger cpu times or the numerical results         
do not influence the conclusions drawn in this study. In Table 6, we present the 
average number of bonds broken at the peak load and at failure per lattice (triangular) 
configuration. It should also be noted that in Table 6, we were able to perform 
emsemble averaging over many number of configurations because we were able to run these simulations 
simultaneously on many number of {\it Eagle} 375 MHz Power3-II processors. 

\par
\begin{figure}[hbtp]
\centerline{\includegraphics[width=12cm]{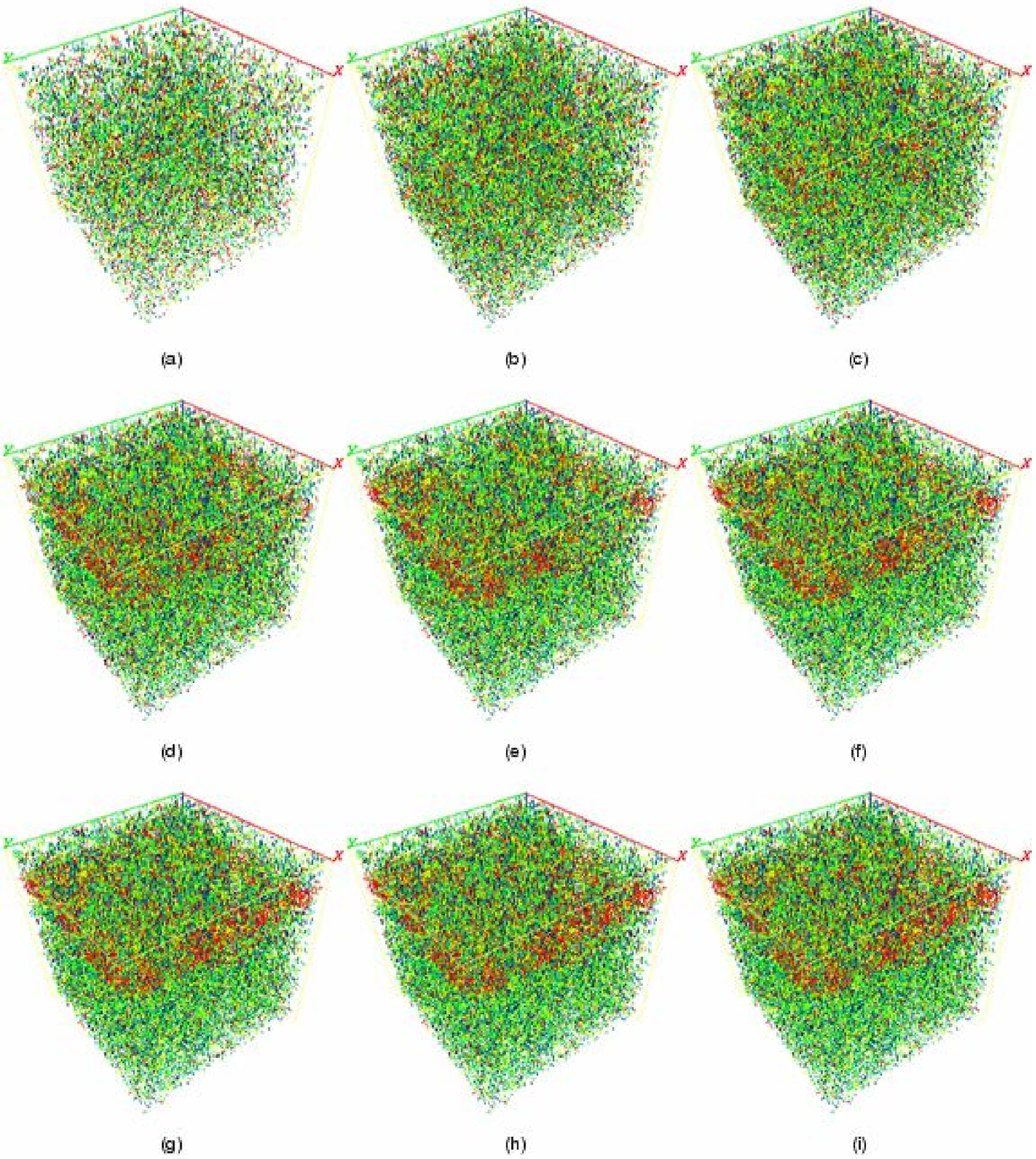}}
\caption{Snapshots of damage in a typical cubic lattice system of size $L = 48$. Number of broken bonds at the peak load and at failure are 48904 and 54744, respectively. (a)-(i) represent the snapshots of damage after breaking $n_b$ number of bonds. The coloring scheme is such that in each snapshot, the bonds broken in the early stages are colored blue, then green, followed by yellow, and finally the last stage of broken bonds are colored red. (a) $n_b = 20000$ (b) $n_b = 40000$ (c) $n_b = 48904$ (peak load) (d) $n_b = 51000$ (e) $n_b = 52500$ (f) $n_b = 53500$ (g) $n_b = 54000$ (h) $n_b = 54500$ (i) $n_b = 54744$ (failure)}
\label{fig1a}
\end{figure}
\begin{figure}[hbtp]
\centerline{\includegraphics[width=12cm]{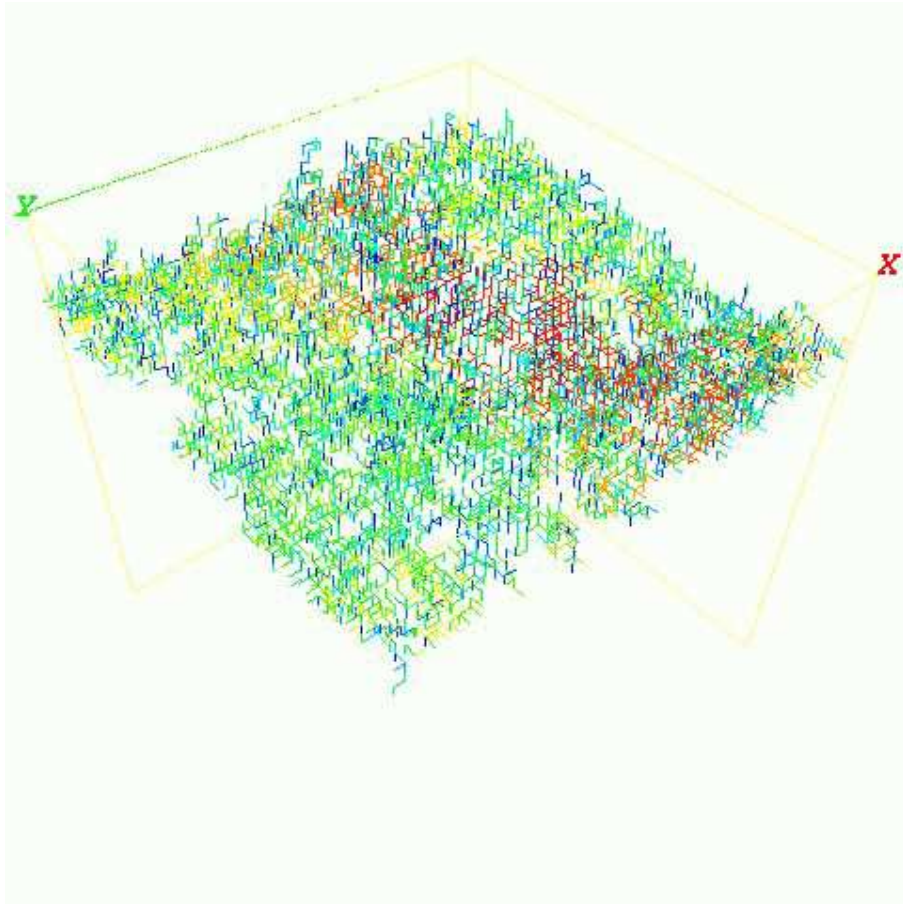}}
\caption{Spanning cluster in a typical cubic lattice system of size $L = 48$. The coloring scheme is such that the bonds broken in the early stages are colored blue, then green, followed by yellow, and finally the last stage of broken bonds are colored red}
\label{fig2a}
\end{figure}

In addition to the above presented simulations on two-dimensional (2D) triangular lattices, we have 
also carried out simulations on three-dimensional (3D) cubic lattice networks to investigate the 
efficiency of block circulant PCG solvers in 3D simulations. 
Figure \ref{fig1a} presents the snapshots of progressive damage
evolution for the case of a broadly distributed random thresholds model problem
in a cubic lattice system of size $L = 48$. The spanning cluster is shown in Fig. \ref{fig2a}.
Tables 7-10 present the cpu and wall-clock 
times taken on a single processor of {\it Cheetah} 
for simulating one-configuration using the block circulant, {\it optimal} circulant, 
un-preconditioned, and the incomplete Cholesky iterative solvers, respectively. 
It should be noted that for large 3D lattice systems (e.g., $L = 32$), 
the performance of incomplete Cholesky preconditioner (see Table 10) is similar to that of 
block-circulant preconditioner (see Table 7), even though the performance of  
incomplete Cholesky preconditioner is far more superior in the case of 2D lattice simulations. 
The memory limitations severely restricted the use of sparse direct solvers for simulating 
large 3D lattice systems, and hence the results corresponding to the direct solver for 3D lattice systems 
are not presented. In the case of 
block circulant PCG, we once again partition the matrix ${\bf A}$ into 
$L$-by-$L$ blocks of size $(L+1)^2 \times (L+1)^2$ matrices. It should be noted that in order to 
get maximum efficiency using the block circulant PCG solver, it is possible to further partition each of the 
$(L+1)^2 \times (L+1)^2$ matrix blocks into $(L+1) \times (L+1)$ blocks of matrices of size 
$(L+1) \times (L+1)$. However, the results presented in this study do not perform such 
nested block circulant precondioning. Table 11 presents the average number of bonds broken 
at the peak load and at failure per lattice configuration.

\par

\section{Conclusions}
The main focus of the current paper is on developing an efficient algorithm 
based on iterative solvers for simulating large 3D fuse networks. Although the 
sparse direct solvers presented in \cite{nukalajpamg} achieve superior 
performance over iterative solvers in 2D lattice systems, the available 
random access memory poses a severe constraint over the usage of sparse direct solvers for large 
3D lattice systems due to the amount of {\it fill-in} during sparse Cholesky factorization. 
In this regard, the block-circulant preconditioner presented in the current paper is an effort toward 
efficiently solving large 3D fuse networks. 
\par
Based on the numerical simulation results presented in Tables 1-5 (2D) and Tables 7-10 (3D) for 
random threshold fuse model networks, it is clear that the block circulant 
preconditioned CG is superior to the {\it optimal} circulant preconditioned PCG solver, 
which in turn is superior to the Fourier accelerated PCG solvers. Furthermore, 
in the case of large 3D lattice systems, the block-circulant preconditioner 
exhibits superior performance (for system sizes $L > 32$) over the sparse direct solvers and 
the related incomplete Cholesky preconditioned CG solvers.  

\par
In addition, during the CG iterative solution, the preconditioned system using the 
block-circulant preconditioner is trivially parallel, and hence a parallel 
implementation of the block-circulant precondioner can be employed to further 
speed up the solution of large 3D lattice systems. This allowed us to consider 
larger 3D lattice simulations, which will be a subject of future publication. 

\par
\vskip 1.00em%
\noindent
{\bf Acknowledgment} \\
This research is sponsored by the Mathematical, Information and Computational Sciences
Division, Office of Advanced Scientific Computing Research, U.S. Department of Energy under
contract number DE-AC05-00OR22725 with UT-Battelle, LLC. The first author wishes to thank
Ed F. D'Azevedo for many helpful discussions and excellent suggestions.

\newpage

\bibliography{bfgs_bfft}
\bibliographystyle{unsrt}

\newpage

\newpage

\newpage

\begin{table}[hbtp]
  \leavevmode
  \begin{center}
  \caption{Block Circulant PCG: 2D Triangular Lattice}
  \vspace*{1ex}
  \begin{tabular}{|c|c|c|c|c|}\hline
  Size  & CPU(sec) & Wall(sec) & Iterations & $N_{config}$ \\
  \hline
 32 & 10.00 & 10.68 & 11597 & 20000 \\
 64 & 135.9 & 139.8 & 41207 & 1600 \\
128 & 2818 & 2846 & 147510 & 192 \\
256 & 94717 & 96500 & & 32 \\
  \hline
  \end{tabular}
  \label{table1}
  \end{center}
\end{table}

\begin{table}[hbtp]
  \leavevmode
  \begin{center}
  \caption{{\it Optimal} Circulant PCG: 2D Triangular Lattice}
  \vspace*{1ex}
  \begin{tabular}{|c|c|c|c|c|}\hline
  Size  & CPU(sec) & Wall(sec) & Iterations & $N_{config}$ \\
  \hline
 32 & 11.66 & 12.26 & 25469 & 20000 \\
 64 & 173.6 & 178.8 & 120570 & 1600 \\
128 & 7473 & 7725 & 622140 & 128 \\
  \hline
  \end{tabular}
  \label{table2}
  \end{center}
\end{table}

\begin{table}[hbtp]
  \leavevmode
  \begin{center}
  \caption{Un-preconditioned CG: 2D Triangular Lattice}
  \vspace*{1ex}
  \begin{tabular}{|c|c|c|c|c|}\hline
  Size  & CPU(sec) & Wall(sec) & Iterations & $N_{config}$ \\
  \hline
 32 & 7.667 & 8.016 & 66254 & 20000 \\
 64 & 203.5 & 205.7 & 405510 & 1600 \\
  \hline
  \end{tabular}
  \label{table3}
  \end{center}
\end{table}

\begin{table}[hbtp]
  \leavevmode
  \begin{center}
  \caption{Incomplete Cholesky PCG: 2D Triangular Lattice}
  \vspace*{1ex}
  \begin{tabular}{|c|c|c|c|c|}\hline
  Size  & CPU(sec) & Wall(sec) & Iterations & $N_{config}$ \\
  \hline
 32 & 2.831 & 3.008 & 5857 & 20000 \\
 64 & 62.15 & 65.61 & 29496 & 4000 \\
128 & 1391 & 1430 & 148170 & 320 \\
  \hline
  \end{tabular}
  \label{table4}
  \end{center}
\end{table}

\begin{table}[hbtp]
  \leavevmode
  \begin{center}
  \caption{Computational cost associated with solver type A of Ref. \cite{nukalajpamg}}
  \vspace*{1ex}
  \begin{tabular}{|c|c|c|c|}\hline
  Size  & CPU(sec) & Wall(sec) & $N_{config}$ \\
  \hline
 32 & 0.592 & 0.687 & 20000 \\
 64 & 10.72 & 11.26 & 4000 \\
128 & 212.2 & 214.9 & 800 \\
256 & 5647 & 5662 & 96 \\
512 & 93779 & 96515 & 16 \\
  \hline
  \end{tabular}
  \label{table4a}
  \end{center}
\end{table}

\begin{table}[hbtp]
  \leavevmode
  \begin{center}
  \caption{Number of broken bonds at peak and at failure}
  \vspace*{1ex}
  \begin{tabular}{|c|c|c|c|c|c|c|c|}\hline
  L  & $N_{config}$ & \multicolumn{4}{c|}{Triangular} \\\cline{3-6}
     & & $n_p$ (mean) & $n_p$ (std) & $n_f$ (mean) & $n_f$ (std) \\
  \hline
  4 & 50000 & 13 & 3 & 19 & 3 \\
  8 & 50000 & 41 & 8 & 54 & 7 \\
 16 & 50000 & 134 & 19 & 168 & 18 \\
 24 & 50000 & 276 & 32 & 335 & 31 \\
 32 & 50000 & 465 & 48 & 554 & 46 \\
 64 & 50000 & 1662 & 130 & 1911 & 121 \\
128 & 12000 & 6068 & 386 & 6766 & 349 \\
256 & 1200 & 22572 & 1151 & 24474 & 1046 \\
  \hline
  \end{tabular}
  \label{table4b}
  \end{center}
\end{table}

\begin{table}[hbtp]
  \leavevmode
  \begin{center}
  \caption{Block Circulant PCG: 3D Cubic Lattice}
  \vspace*{1ex}
  \begin{tabular}{|c|c|c|c|c|}\hline
  Size  & CPU(sec) & Wall(sec) & Iterations & $N_{config}$ \\
  \hline
10 & 16.54 & 16.99 & 16168 & 40000 \\
16 & 304.6 & 308.5 & 58756 & 1920 \\
24 & 2154 & 2216 & 180204 & 256 \\
32 & 12716 & 12937 & 403459 & 128 \\
48 & 130522 & 133063 & 1253331 & 32 \\
  \hline
  \end{tabular}
  \label{table5}
  \end{center}
\end{table}

\begin{table}[hbtp]
  \leavevmode
  \begin{center}
  \caption{{\it Optimal} Circulant PCG: 3D Cubic Lattice}
  \vspace*{1ex}
  \begin{tabular}{|c|c|c|c|c|}\hline
  Size  & CPU(sec) & Wall(sec) & Iterations & $N_{config}$ \\
  \hline
10 & 15.71 & 16.10 & 27799 & 40000 \\
16 & 386.6 & 391.1 & 121431 & 1920 \\
24 & 2488 & 2548 & 446831 & 256 \\
32 & 20127 & 20380 & 1142861 & 32 \\
48 & 233887 & 237571 & 4335720 & 32 \\
  \hline
  \end{tabular}
  \label{table6}
  \end{center}
\end{table}

\begin{table}[hbtp]
  \leavevmode
  \begin{center}
  \caption{Un-preconditioned CG: 3D Cubic Lattice}
  \vspace*{1ex}
  \begin{tabular}{|c|c|c|c|c|}\hline
  Size  & CPU(sec) & Wall(sec) & Iterations & $N_{config}$ \\
  \hline
10 & 5.962 & 6.250 & 48417 & 40000 \\
16 & 119.4 & 123.0 & 246072 & 3840 \\
24 & 1923 & 1982 & 1030158 & 256 \\
32 & 16008 & 16206 & 2868193 & 64 \\
  \hline
  \end{tabular}
  \label{table7}
  \end{center}
\end{table}

\begin{table}[hbtp]
  \leavevmode
  \begin{center}
  \caption{Incomplete Cholesky PCG: 3D Cubic Lattice}
  \vspace*{1ex}
  \begin{tabular}{|c|c|c|c|c|}\hline
  Size  & CPU(sec) & Wall(sec) & Iterations & $N_{config}$ \\
  \hline
10 & 5.027 & 5.262 & 8236 & 40000 \\
16 & 118.1 & 122.3 & 42517 & 3840 \\
24 & 1659 & 1705 & 152800 & 512 \\
32 & 12091 & 12366 & 422113 & 64 \\
  \hline
  \end{tabular}
  \label{table8}
  \end{center}
\end{table}

\begin{table}[hbtp]
  \leavevmode
  \begin{center}
  \caption{Number of broken bonds at peak and at failure}
  \vspace*{1ex}
  \begin{tabular}{|c|c|c|c|c|c|c|c|}\hline
  L  & $N_{config}$ & \multicolumn{4}{c|}{Cubic} \\\cline{3-6}
     & & $n_p$ (mean) & $n_p$ (std) & $n_f$ (mean) & $n_f$ (std) \\
  \hline
 10 & 40000 & 563 & 57 & 726 & 59 \\
 16 & 3840 & 2108 & 147 & 2572 & 152 \\
 24 & 512 & 6692 & 354 & 7882 & 337 \\
 32 & 128 & 15329 & 705 & 17691 & 649 \\
 48 & 32 & 49495 & 1582 & 55768 & 1523 \\
  \hline
  \end{tabular}
  \label{table10}
  \end{center}
\end{table}

\end{document}